\documentclass[11pt]{article}
\usepackage{latexsym}
\usepackage{graphicx}
\begin{document}
\title{Spontaneous breaking of residual gauge symmetries on the lattice}
\author{Michael Grady\\
Department of Physics\\ State University of New York at Fredonia\\
Fredonia NY 14063 USA}
\date{\today}
\maketitle
\thispagestyle{empty}
\begin{abstract}
Lattice gauge theories are considered with a partial axial gauge fixing 
along one direction only.  This leaves a residual gauge symmetry that is still
local in three directions but now global in one.  It is found that this
$N^{d-1}$ fold symmetry (on an $N^d$ lattice) breaks spontaneously at weak 
coupling with the gauge field elements on links averaged over 1-d chains along the gauge-fixing
direction as order parameters. This phase transition is observed with Monte-Carlo
simulations for both 3-d Z2 and 4-d SU(2) pure gauge theories and appears to be
coincident with the deconfinement transition. This work calls into question the 
equivalence of different gauges in certain circumstances.
\end{abstract}
\section{Introduction}
It is well known that Elitzur's theorem\cite{elitzur} prevents the spontaneous 
breaking of a local gauge symmetry.  However, once the gauge is fixed with a 
suitable gauge-fixing
term in the Lagrangian, then the remaining global gauge symmetry can
be spontaneously broken, such as by the Higgs field in the standard model. In this letter
the possibility that the global symmetry (or partially-global gauge symmetry 
that results from
partial gauge fixing) might already be spontaneously broken
by the gauge fields themselves is explored.
This is known to take place in continuum quantum electrodynamics, where the spontaneous
breaking of a residual gauge symmetry left over after formulating in the Lorentz gauge
has been shown to account for the masslessness of photons. In this picture the photons 
are seen as
Goldstone bosons associated with the spontaneously broken symmetries\cite{sbqed}. 
The situation 
for non-abelian continuum
theories is not as clear.

Earlier papers explored the sub-maximal axial gauge, defined by omitting 
the last line of links from the usual maximal axial gauge-fixing tree, leaving an N-fold gauge symmetry
still local in one direction\cite{lsb,z2h}. Here it was found that these residual gauge symmetries 
break spontaneously at weak coupling
and are unbroken at strong coupling.  These regions are separated by a phase transition which
appears to be coincident with the deconfinement transition in both the 3-d Z2 and
the 4-d SU(2) theories.  The order parameters for these transitions are the N-th direction
pointing links (gauge elements) averaged over N-1 dimensional layers perpendicular to the Nth direction.
These systems have a large vacuum degeneracy since each layer can lie in a different
broken-symmetry direction.  This can have an effect on the critical behavior at the 
phase transition because the sudden loss in ergodicity across the phase transition will
result in a change in the entropy (because some sectors are no longer 
being visited, the number of
microstates changes). In a more or less restrictive gauge the amount of entropy change 
at the phase transition will be different
because the vacuum will have a different multiplicity change at the
transition (see below).  
The usual argument that 
gauge fixing is irrelevant because it multiplies the partition function
by the same ``overcounting" factor breaks down
because this factor is different on the two sides of the transition. If a symmetry is 
spontaneously broken the multiple vacua are 
{\em already} only being single-counted due to normal loss
of ergodicity - the other vacuum sectors are excluded because they cannot be tunneled to.
Choosing a more-restrictive gauge 
may select a particular vacuum, but does not change the size of the accessible ensemble. 
However, in the 
unbroken confining phase, which is ergodic, gauge fixing
does change multiplicities by restricting the space of available states. The more-restrictive 
gauges will change this factor by a larger amount then the less-restrictive ones. Therefore it would appear
that different gauges will have different results at such a phase transition.

In a previous paper it was shown that 
the Fradkin-Shenker proof\cite{fs} of the lack of a phase transition between
Higgs and confinement phases in gauge-Higgs theories is only valid in fully-fixed
unitary gauge\cite{z2h}. When the gauge condition is relaxed to allow for the symmetry to 
be local in one direction then these regions {\em are} separated 
by a symmetry-breaking phase transition.
The layered gauge symmetry is spontaneously broken in the Higgs phase and unbroken
in the confinement phase. Thus the critical behavior is drastically different
in the different gauges. By locking down the gauge symmetry completely, the unitary 
gauge explicitly breaks all of the gauge symmetries 
allowing no room for spontaneous breaking
to occur.
   
It is interesting to consider how far one can go in the program of loosening the 
gauge fixing so as to allow for a greater multiplicity of gauge symmetries and still
see spontaneous breaking.  In this paper it is shown that even if the gauge fixing 
is limited to a single direction, leaving the gauge symmetry still local in N-1 directions
and global in only one, symmetry breaking still takes place in both the discrete
and continuous cases. This is somewhat surprising at first, in that the order parameters
are now the average of gauge-links along only one-dimensional chains lying in the gauge-fixing 
direction.  Spontaneous symmetry breaking is known not to occur in one-dimensional systems
at non-zero temperature. However, these chains interact with other chains 
through the gauge couplings. The action is still N-dimensional, so the arguments 
preventing 1-d symmetry breaking are not valid.
There are now $N^{d-1}$ separate symmetries breaking, resulting 
in an even much greater degree of
vacuum degeneracy.  If one's intent is to find a gauge-fixed system that has the 
same behavior as the unfixed system, then it is important not to fix beyond the level
one needs to invalidate Elitzur's theorem to allow for 
symmetry breaking. In other words, one should
not fix symmetries that are already spontaneously broken, because that will introduce
different counting weights on the two sides of the transition than exist in the unfixed case.
Since Elitzur's theorem requires an infinite number of fields to be involved in
a symmetry transformation for it to be able to break spontaneously, it would appear we have 
essentially reached this limit with the infinite chain. One could perhaps leave out
one more gauge-fixing on a single link in each chain leaving two semi-infinite chains,
but probably this
is too small an addition to the vacuum degeneracy to affect anything in the infinite
volume limit.  Therefore the results in this minimal axial gauge involving only links
in a single direction should be comparable to those of the unfixed case.
The reason for fixing the gauge at all is to allow the transition to be visible through the 
definition of the abovementioned order parameters. The same transitions are in-principle
observable in the unfixed case by using non-local order parameters, as will be detailed 
below. It would appear, therefore that our previous unmasking of the layered gauge
symmetry breaking did not go far enough. The number of broken symmetries is much larger than 
$N$-fold -- it is $N^{d-1}$ fold. The picture that emerges for the weak-coupling phase
is one of intersecting bundles of 
spin chains locked into a high-degeneracy broken-symmetry pattern.

\section{Monte Carlo simulations}

Simulations were performed in both the 3-d Z2 pure lattice gauge theory (i.e.\ with
no matter fields) and also the SU(2) pure-gauge theory in four dimensions on
symmetric lattices with periodic boundary conditions. The abovementioned
minimal axial gauge was implemented by setting all links in the 1-direction equal to unity
except forward-pointing links attached to sites with $x_1 =0$.  The remaining links in 
the 1-direction cannot be set to unity by a gauge transformation if one is
using periodic boundary conditions.  These remaining 1-links 
are gauge-covariant and are equal to the Polyakov loops in that direction.
Normally, one completes the gauge fixing by building a tree of fixed links in other directions
in the $x_1 =0$ (hyper)surface. However, we leave these other links unfixed.  This leaves
a large residual gauge group unfixed. Any gauge transformation which is independent of $x_1$
is still allowed.  Thus the residual symmetry is global in the 1-direction but still local
in all of the other directions.

\subsection{Z2 in three dimensions}
\begin{figure}[ht]
                      \includegraphics[width=2.5in]{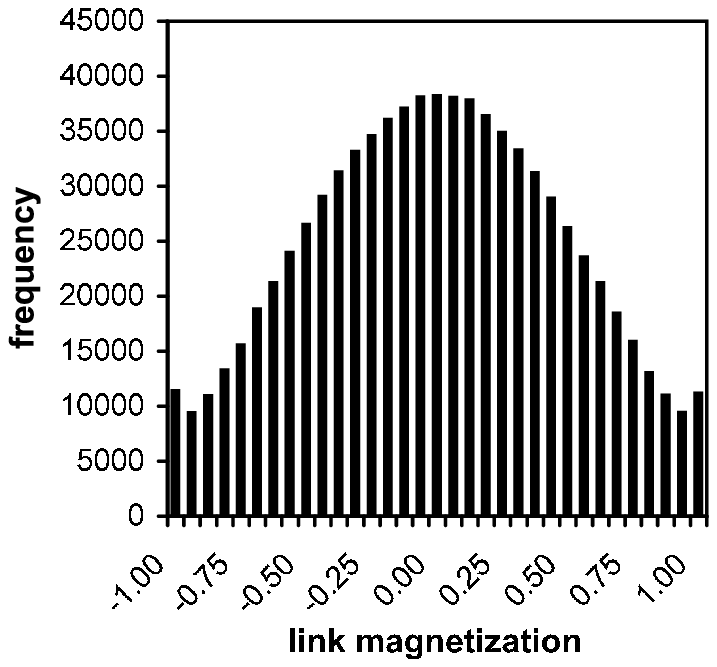}
                      \includegraphics[width=2.5in]{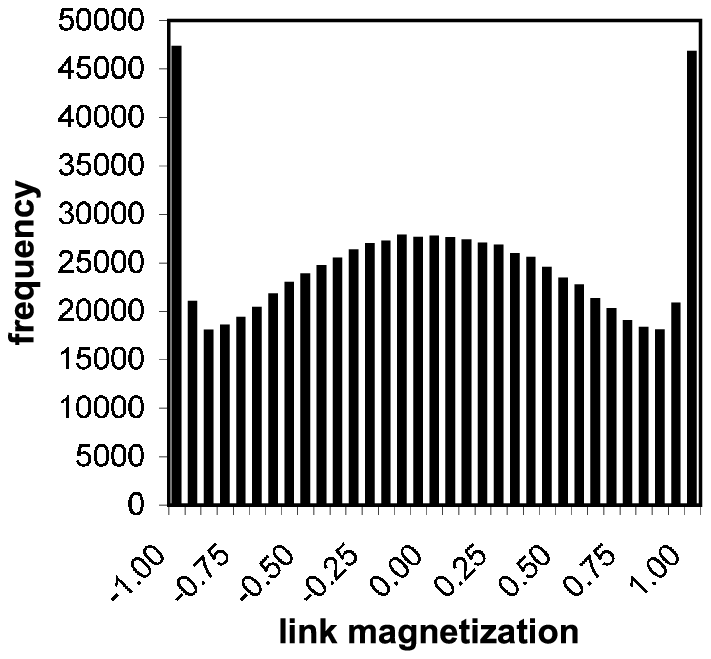}
                      \includegraphics[width=2.5in]{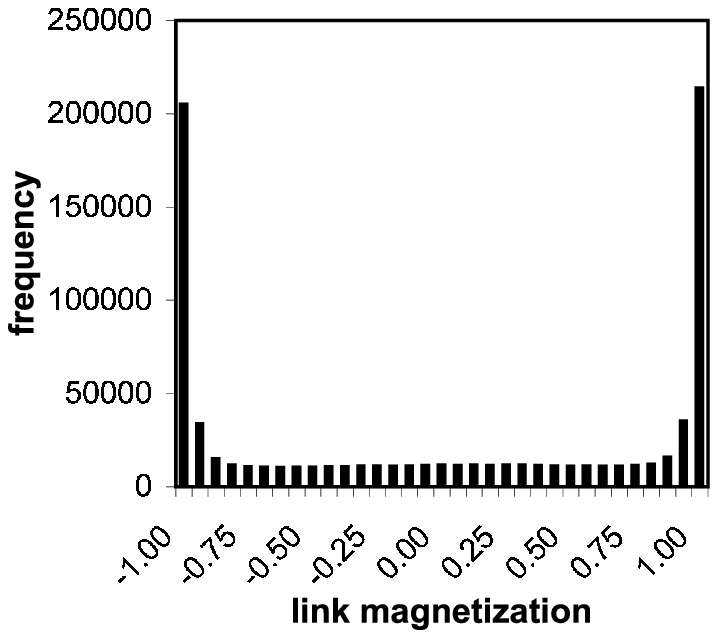}
            \caption{Distributions of average link magnetizations on 1-d chains
for the 3-d Z2 lattice gauge theory on a $32^3$ lattice for 
couplings $\beta=$ (a)~0.69, (b)~0.73, and (c)~0.76. The infinite-lattice critical point
is expected to lie at 0.7613.}
          \label{fig1}
       \end{figure}

\begin{figure}[ht]
                      \includegraphics[width=2.5in]{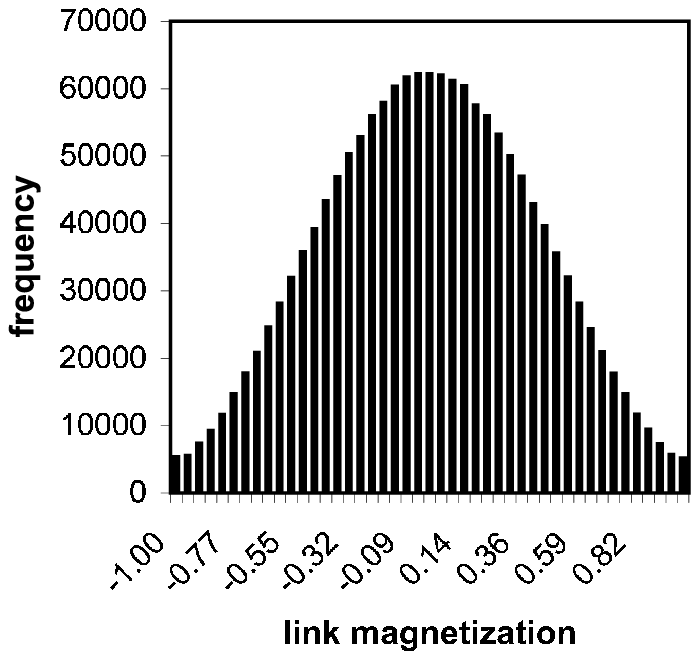}
                      \includegraphics[width=2.5in]{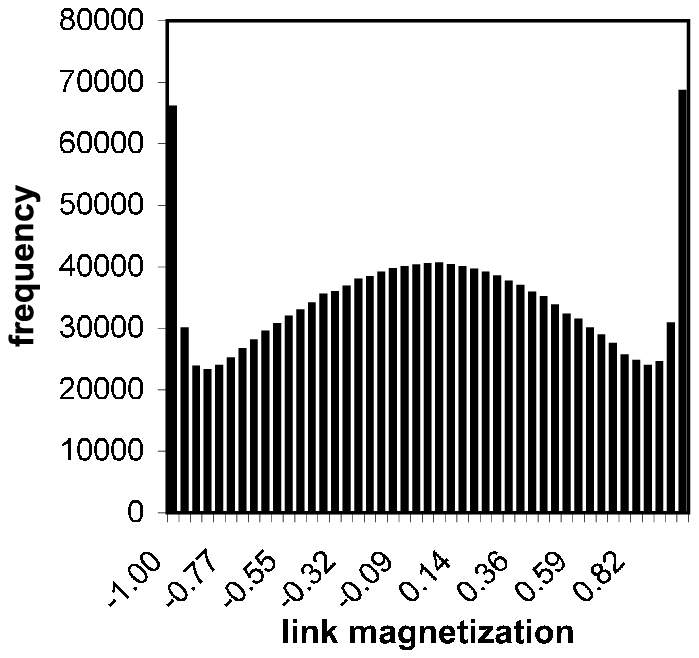}
                      \includegraphics[width=2.5in]{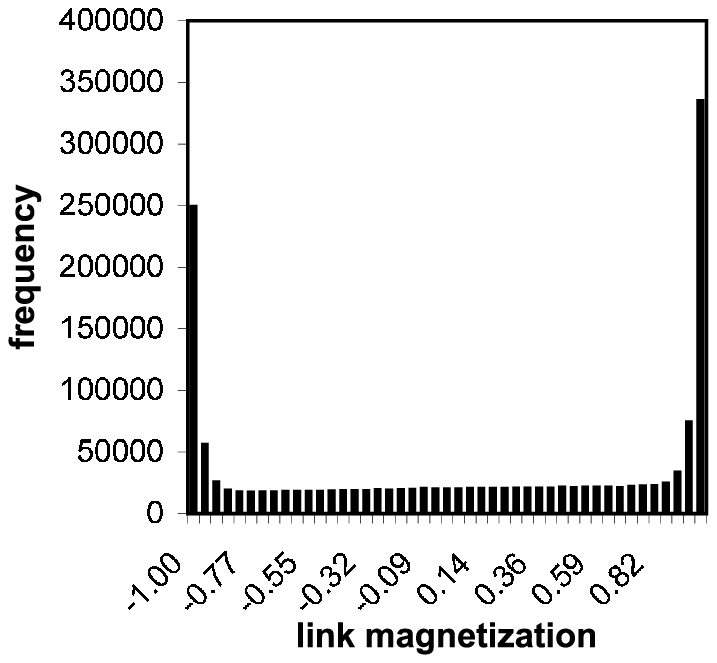}
            \caption{Same as Fig.~\ref{fig1} except on a $44^3$ lattice at 
couplings $\beta=$ (a)~0.69, (b)~0.74, and (c)~0.76.}
          \label{fig2}
       \end{figure}

For the 3-d Z2 lattice gauge theory, Figs.~\ref{fig1},\ref{fig2} shows the 
histograms of gauge links lying in other directions averaged along chains
in the $x_1$
direction. Each lattice supplies $(d-1)*N^{(d-1)}$
independent samples for this distribution, leading to high statistics in a relatively
small sample. Each separate residual gauge transformation that
affects a chain of sites in the $x_1$ direction will transform $2(d-1)$ attached link-chains.
Therefore if the chain magnetizations acquire expectation values 
then the attached gauge symmetries will be spontaneously broken.
Monte Carlo simulations were run for 20,000 sweeps, after 10,000 equilibration sweeps, with
measurements taken every 50 sweeps. The Z2 theory has a deconfinement phase transition
at $\beta = 0.7613$ on the infinite lattice.  This is known because it is dual to the 
three-dimensional Ising model\cite{wegner},
and has been confirmed with Monte-Carlo simulations\cite{stack}. A clear symmetry-breaking is
shown in the histograms, with distributions peaked at zero on the confining side
of the transition and away from zero in the deconfined phase. In the intermediate
region three-peaked histograms with one peak at zero and two symmetrically away from
zero give an apparent indication of a first-order phase transition. The modest deepening
of valleys when one moves from the $32^3$ to $44^3$ lattice is also indicative
of a first-order transition.  This seems puzzling, because duality with the 3-d Ising
model and previous Monte-carlo simulations would indicate a second-order transition.
However, the duality transformation actually relates the 3-d Ising model to the Ising gauge
theory in a fully-fixed maximal-tree axial gauge.  As seen above, if parts of the
gauge symmetry itself break spontaneously, then one actually has good reason to
expect different critical behavior in the fully-fixed gauge than in the minimal
axial gauge being explored here. In a similar vein, the Higgs-confinement transition, seen  
in the sub-maximal axial gauge in Ref.~\cite{z2h}, is completely absent in the unitary gauge
employed in the Fradkin-Shenkar proof of phase continuity 
between Higgs and confinement regions.  Thus phase transitions of this sort do
depend on the gauge being used, so different orders in different 
gauges are not out of the question. Further study, such as a full finite-size scaling
analysis will be needed to verify the order of the transition. The primary purpose 
here is simply to demonstrate the existence of the symmetry breaking coincident
with the deconfinement transition and the usefulness 
of these order parameters. Indeed, because of the dual connection to the Ising model,
this must be a symmetry-breaking phase transition. However, no broken symmetry or 
symmetry breaking order parameter
had been previously identified for the gauge theory. Therefore it is not surprising
to find a hidden symmetry breaking here. 

This transition was briefly studied in the sub-maximal
axial gauge previously\cite{z2h}. A transition was verified, 
but the simulations were plagued with
equilibration difficulties. Even after several hundred thousand sweeps there were
discrepancies between hot and cold starts when well into the symmetry-broken region. 
In the present minimal axial gauge no
problems with equilibration have been seen. This is probably because now only $2(d-1)N$ links are
involved with each symmetry transformation, as opposed to $2(d-1)N^3$. Needless to
say, it is much easier to flip the former than the latter number of links.
The histograms for cold-start
simulations in the broken phase 
with only 10,000 equilibration sweeps are already nearly left-right symmetrical, showing the 
ease of tunneling between vacua (the cold start had all links initially set to unity).

\subsection{SU(2) in four dimensions}
The 4-d SU(2) theory will first be considered in the first-order region of
the fundamental-adjoint plane, because the position of the bulk transition is 
well defined there through the jump in the plaquette.  Whether or not this
bulk transition is symmetry-breaking is a very important question.  
Indirect evidence from the scaling of the size of the
metastability region with the latent heat strongly suggests it is symmetry breaking\cite{tcp}.
The Landau theory makes a clear distinction between first-order transitions
which are symmetry-breaking and those which are not, leading to a prediction of quadratic
scaling in one case and linear in the other. The data clearly support the case
of a symmetry-breaking transition\cite{tcp}.
If the transition is symmetry-breaking, then the weak coupling side,
which includes the  continuum
limit, has a different symmetry than the strong coupling (confining) side. For
an exact symmetry, these must everywhere be separated by a phase transition. 
The first-order endpoint in the fundamental-adjoint plane would necessarily be
a tricritical point rather than an ordinary liquid-gas type critical point. This means it 
would not be possible to find an analytic path around this point to connect the confining
phase to the continuum limit, as these would be in different symmetry regions.

\begin{figure}[ht]
                      \includegraphics[width=2.5in]{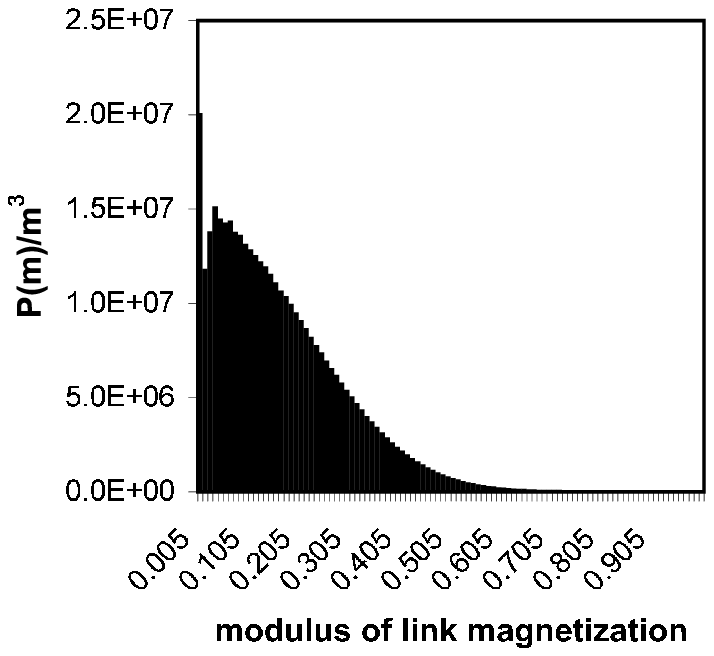}
                      \includegraphics[width=2.5in]{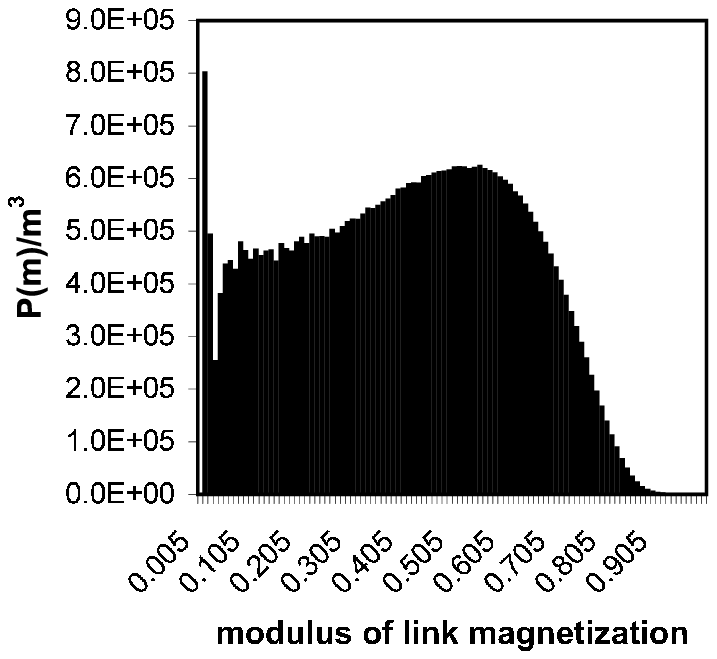}
                      \includegraphics[width=2.5in]{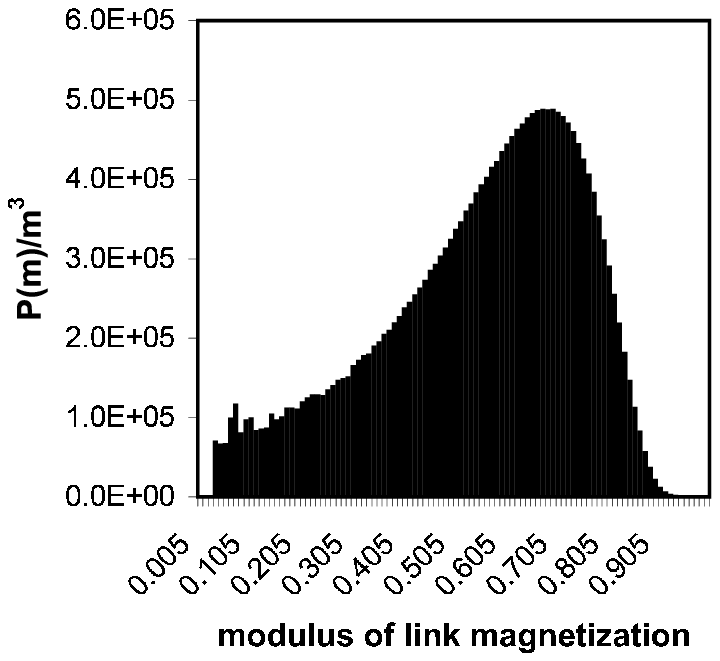}
              \caption{Distributions of the modulus of the average link magnetization
on 1-d chains for the 4-d SU(2) lattice gauge theory in the fundamental-adjoint plane
with $\beta _{adj} = 1.5$ on a $16^4$ lattice. Fundamental couplings are $\beta =$ (a)~1.0,
(b)~1.1, and (c)~1.5. The known first-order transition lies near
$\beta = 1.04$}
          \label{fig3}
       \end{figure}

\begin{figure}[ht]
                      \includegraphics[width=2.5in]{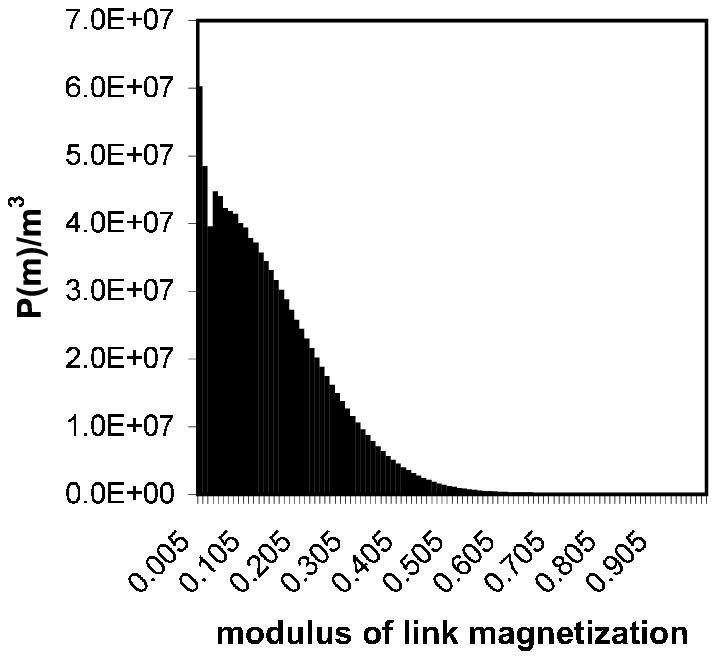}
                      \includegraphics[width=2.5in]{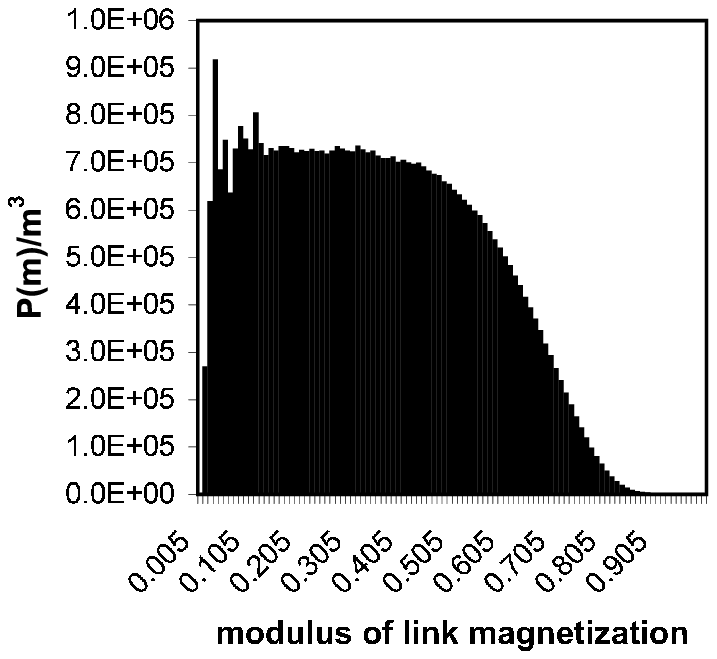}
                      \includegraphics[width=2.5in]{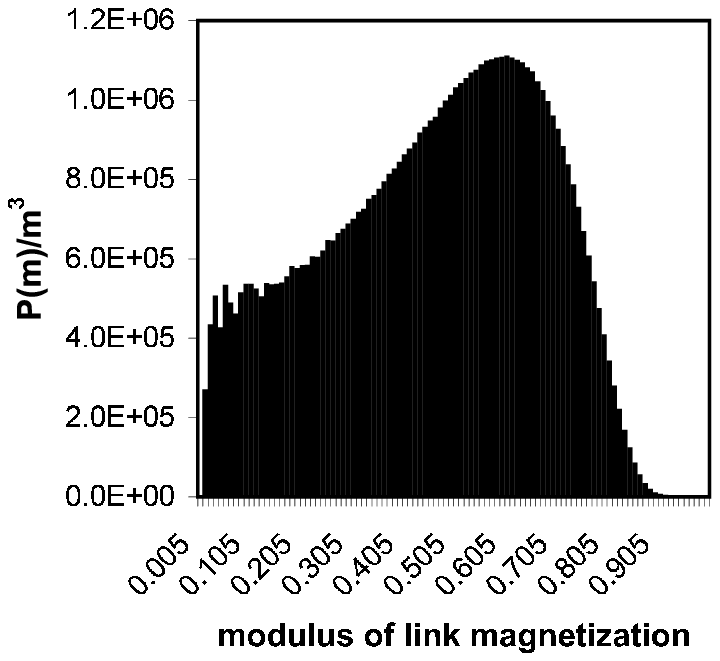}
            \caption{Same as \ref{fig3} except on a $20^4$ lattice.}
          \label{fig4}
       \end{figure}

For this theory the SU(2) links in a given direction are averaged along chains, again lying in the 
1-direction which is the gauge-fixing direction. This average link is, of course, no longer a normalized SU(2)
gauge element. It is reinterpreted as an average spin in a 4-d space, with 
the components being the coefficients of the unit matrix and three Pauli matrices.
Individual SU(2) gauge elements are
unit vectors in this space.  
For this O(4) order parameter one must take 
into account the geometrical factor 
(from solid angle) that biases the distribution toward
larger magnitudes. In the unbroken phase, 
the distribution of magnetization moduli, $m$, is
expected to be a factor of $m^3$ times a 
Gaussian, $\exp (-m^2/2 \sigma _{m} ^2)$. To more easily
see the Gaussian behavior, the probability distribution $P(m)$ is 
obtained by histogramming, and
the quantity $P(m)/m^3$ is plotted. 
Figs.~\ref{fig3},\ref{fig4} show the these distributions 
just below the strong first-order 
transition, which occurs for $\beta _{adj} = 1.5$ at $\beta = 1.04\pm 0.02$\cite{tcp},
and just above it.  Gauge links are first averaged
along each chain in the gauge-fixing direction, and then the modulus is taken.
The value of $m$ for each 
bin is not taken at the center,
but at a value that would produce a flat 
histogram in an $m^3$ distribution, regardless
of bin-size choice. This is 
\begin{equation}
m^{3}_{\rm{bin}}=\frac{1}{4}\frac{(m_{2}^{4}-m_{1}^{4})}{(m_2 - m_1 )} .
\end{equation}
where $m_2$ and $m_1$ are the bin edges. This 
detail affects only the first couple of bins in the 
histograms shown.
At this value of the adjoint coupling the average plaquette jumps about 0.27 at
the transition. One can see from the histograms that a definite symmetry breaking takes
place.  The widening of distributions for the $20^4$ lattice over the $16^4$
seems atypical for a first-order transition, however. This will be discussed below.

\begin{figure}[ht]
                      \includegraphics[width=5in]{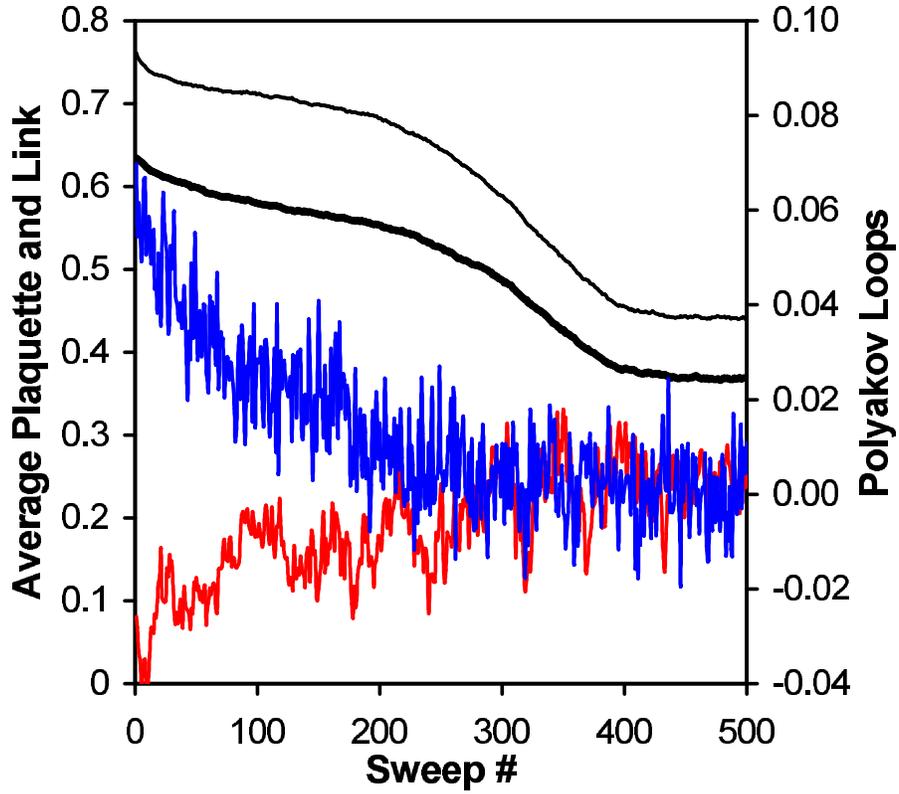}
             \caption{Time history of a quench starting in the ordered phase
of the SU(2) theory near the first-order line in the fundamental-adjoint plane.
The evolution of plaquette (upper thin line), 
average link magnetization (middle thicker line), and average Polyakov loops (lower
two lines -- right scale) are shown. Only two of the four Polyakov loops are shown for clarity. 
The other two
behave similarly.}
          \label{fig5}
       \end{figure}

Fig.~\ref{fig5} shows the time history of a quench. For a lattice equilibrated at
$\beta=1.15$, $\beta _{adj} =1.5$, $\beta$ was suddenly changed to 1.0 at the
beginning of Monte-Carlo time shown in the figure. One can see that the plaquette,
link magnetization, and Polyakov loop all appear to tunnel coincidently. This, together
with the energy scaling evidence given in \cite{tcp}
strongly supports the idea that this is a single integrated  bulk transition 
which is both symmetry-breaking and deconfining. Previously there had been speculation
that a finite-temperature deconfining transition could be lying on top of an
unrelated bulk transition\cite{separate}.  An important point to be made here is that the new order parameter
(link magnetization in chains) is the average of a {\em local quantity}
and thus a bulk order parameter. This
contrasts with the Polyakov loop which is a global order parameter that exists only for
periodic boundary conditions. In addition, when the link magnetization
order parameters acquire  non-zero
expectation values both the residual gauge symmetry {\em and the Polyakov loop symmetry}
are spontaneously broken.
The Polyakov loop symmetry, which multiplies all links in a particular direction in a single
perpendicular hyperlayer by -1, also flips the magnetization of all spin-chains 
made of that-direction links
in that hyperlayer.  Once the Polyakov loop symmetry is broken, there is nothing to protect the Polyakov
loops from gaining an expectation value, which they are seen to do.  Thus the 
residual gauge symmetry
breaking naturally carries with it the Polyakov loop symmetry breaking and thus deconfinement.

\begin{figure}[ht]
                      \includegraphics[width=2.5in]{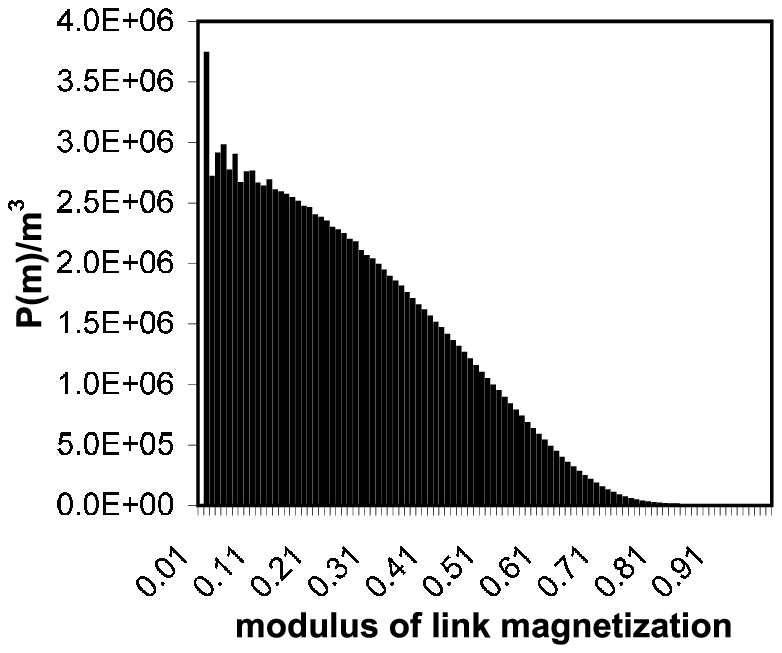}
                      \includegraphics[width=2.5in]{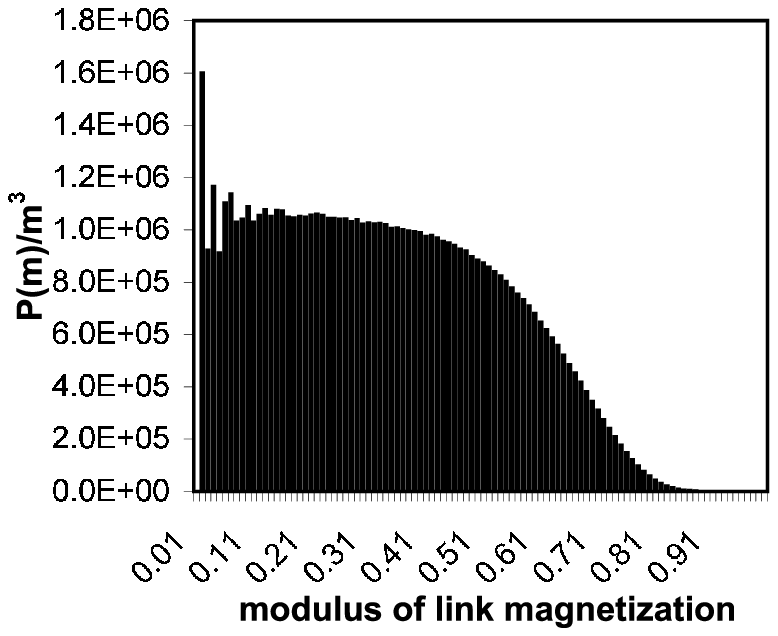}
                      \includegraphics[width=2.5in]{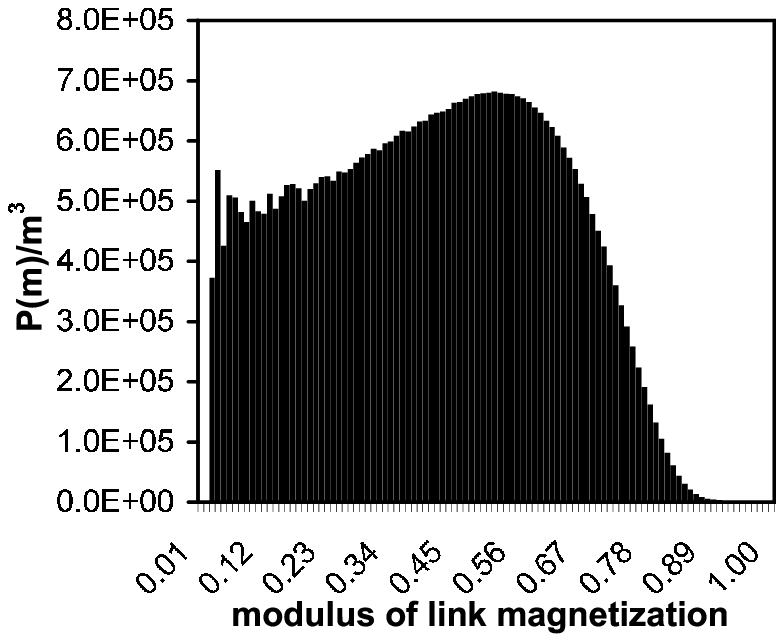}
                      \includegraphics[width=2.5in]{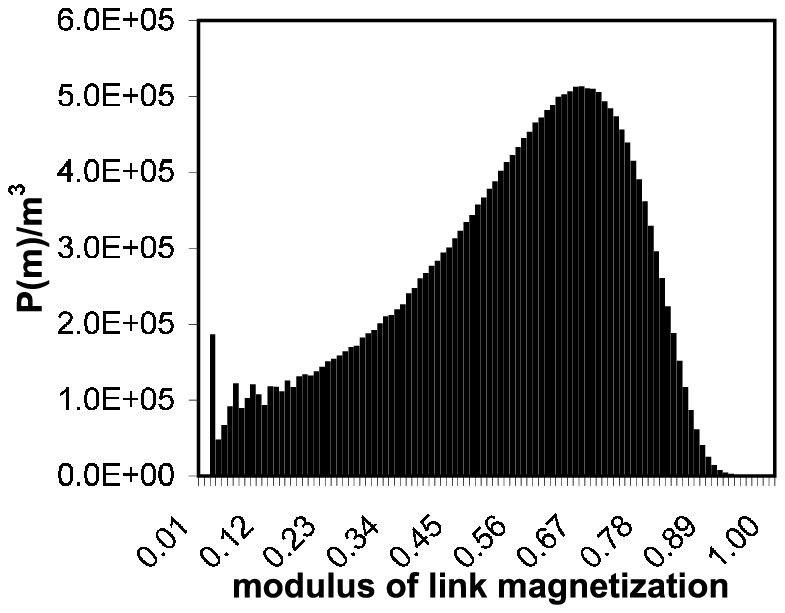}
            \caption{Distributions of the modulus of averge link magnetization
on 1-d chains for the 4-d SU(2) theory on the Wilson axis at (a) $\beta =2.4$,
(b) $\beta =2.8$, (c) $\beta = 3.2$  and (d) $\beta = 4.0$ on a $16^4$ lattice.}
          \label{fig6}
       \end{figure}

\begin{figure}[ht]
                      \includegraphics[width=2.5in]{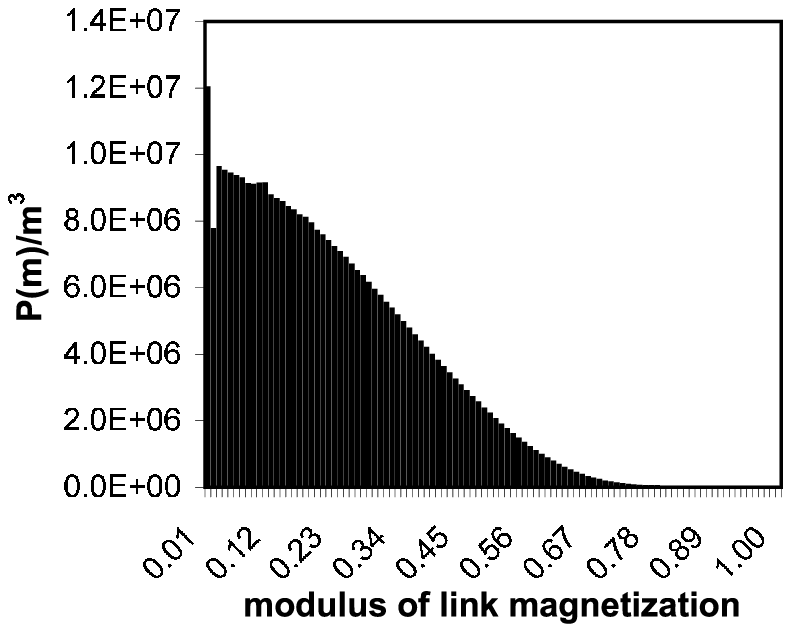}
                      \includegraphics[width=2.5in]{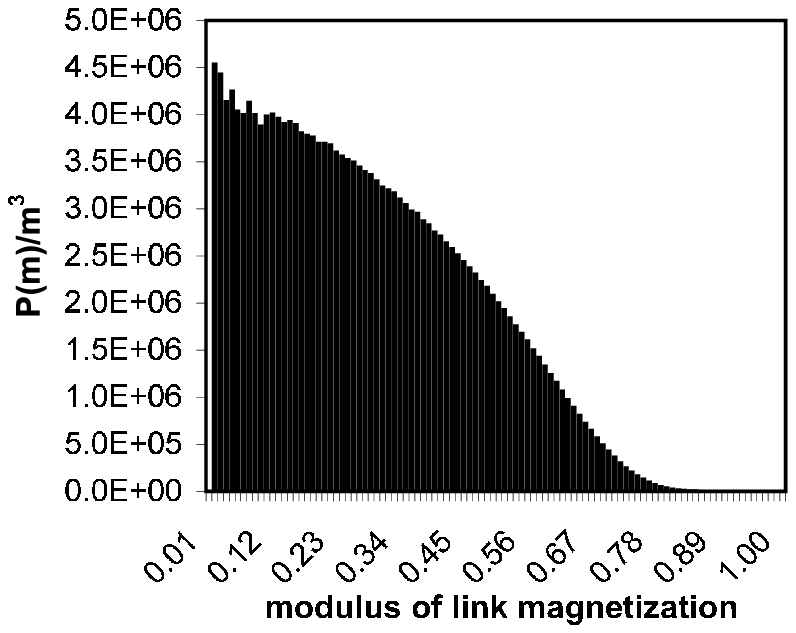}
                      \includegraphics[width=2.5in]{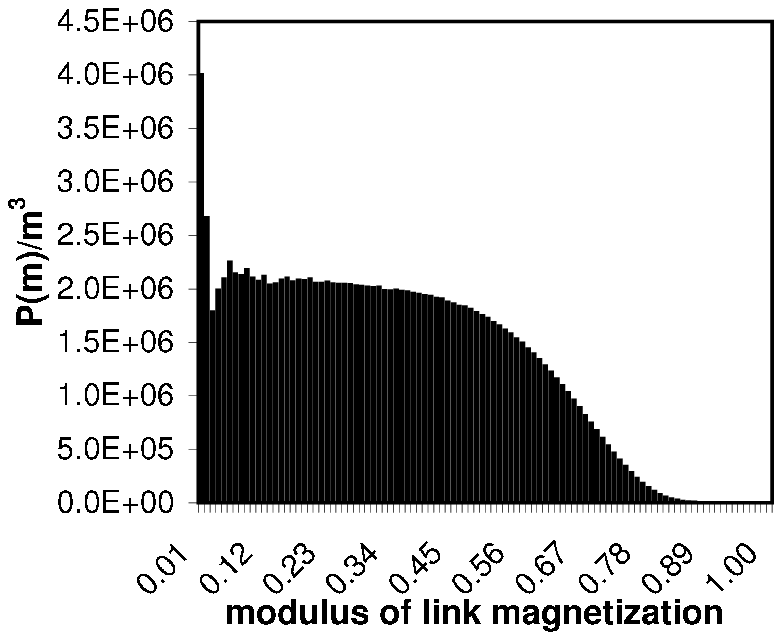}
                      \includegraphics[width=2.5in]{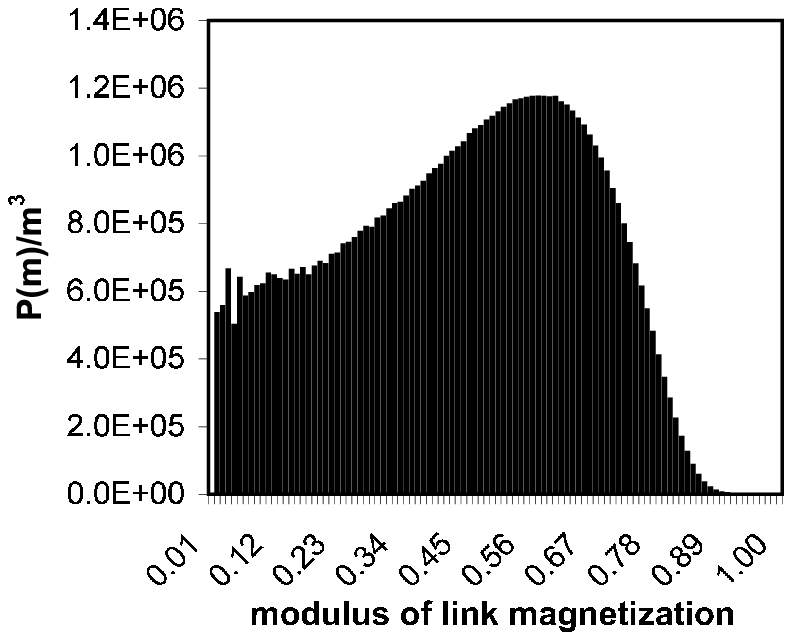}
            \caption{Same as Fig.~\ref{fig6} except on a $20^4$ lattice.}
          \label{fig7}
       \end{figure}

\begin{figure}[ht]
                      \includegraphics[width=5in]{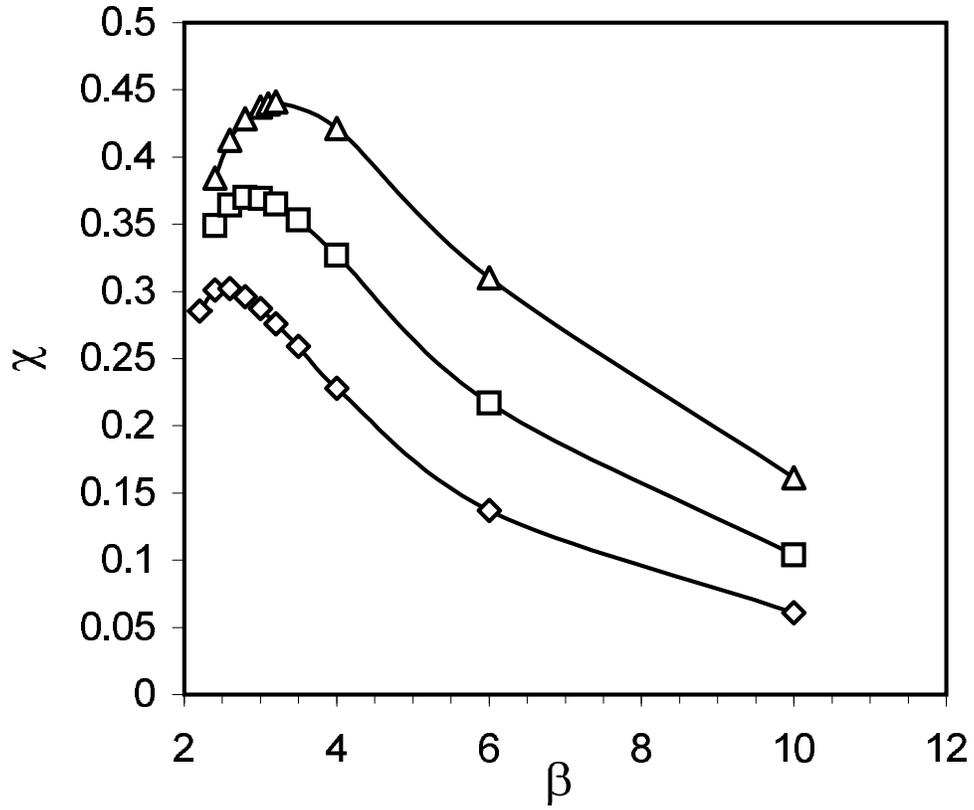}
             \caption{Susceptibility of the link magnetization on 1-d
chains for 4-d SU(2) on the Wilson axis for $16^4$ (diamonds), $20^4$ (squares), 
and $24^4$ (triangles) lattices. Error
bars are about one-fifth the size of plotted points.}
          \label{fig8}
       \end{figure}

Moving on to the Wilson axis, the histograms in Figs.~\ref{fig6},\ref{fig7} 
also show a symmetric phase 
in the confining region,
and a symmetry-broken phase at weak coupling.  
In the intermediate regions a flat-topped distribution is seen, suggestive
of a higher-order transition in this case. The connected susceptibilities for the
link magnetizations averaged on chains are shown in Fig.~\ref{fig8}.
Broad peaks are seen which are definitely growing with lattice size. 
It is not surprising that the transition is broad, because the ``system size"
for each individual order parameter is only $N$, i.e. only $N$ links are being averaged as 
opposed to $N^4$ for an ordinary global symmetry breaking. 

\begin{figure}[ht]
                      \includegraphics[width=2.5in]{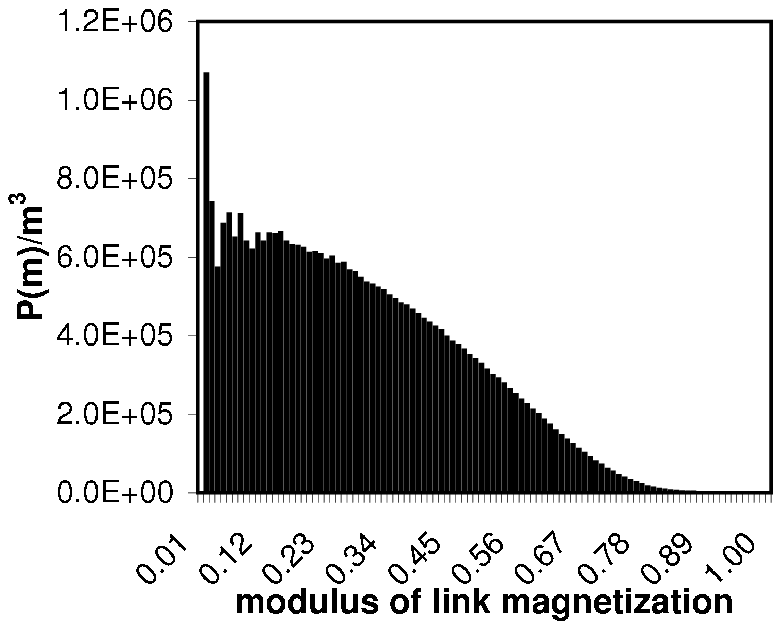}
                      \includegraphics[width=2.5in]{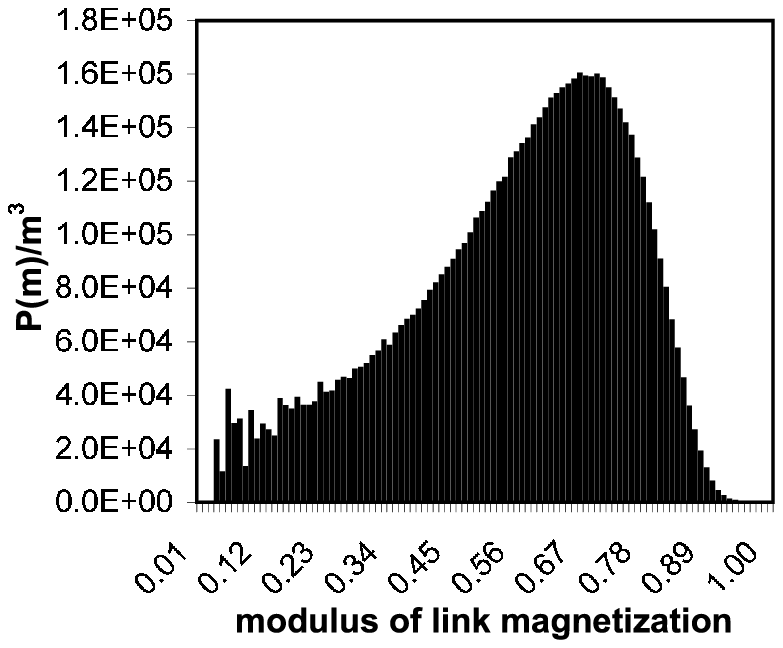}
         \caption{SU(2) Wilson axis runs with open boundary conditions in
all directions on a $16^4$ lattice. (a) $\beta = 2.2$ and (b) $\beta = 3.2$. 
Open boundary condition runs were also performed in the
fundamental-adjoint plane (not shown). These also produced
results similar to those obtained with periodic boundary conditions.
In general variances are larger than with periodic boundary conditions.}
          \label{fig9}
       \end{figure}

\begin{figure}[ht]
                      \includegraphics[width=5in]{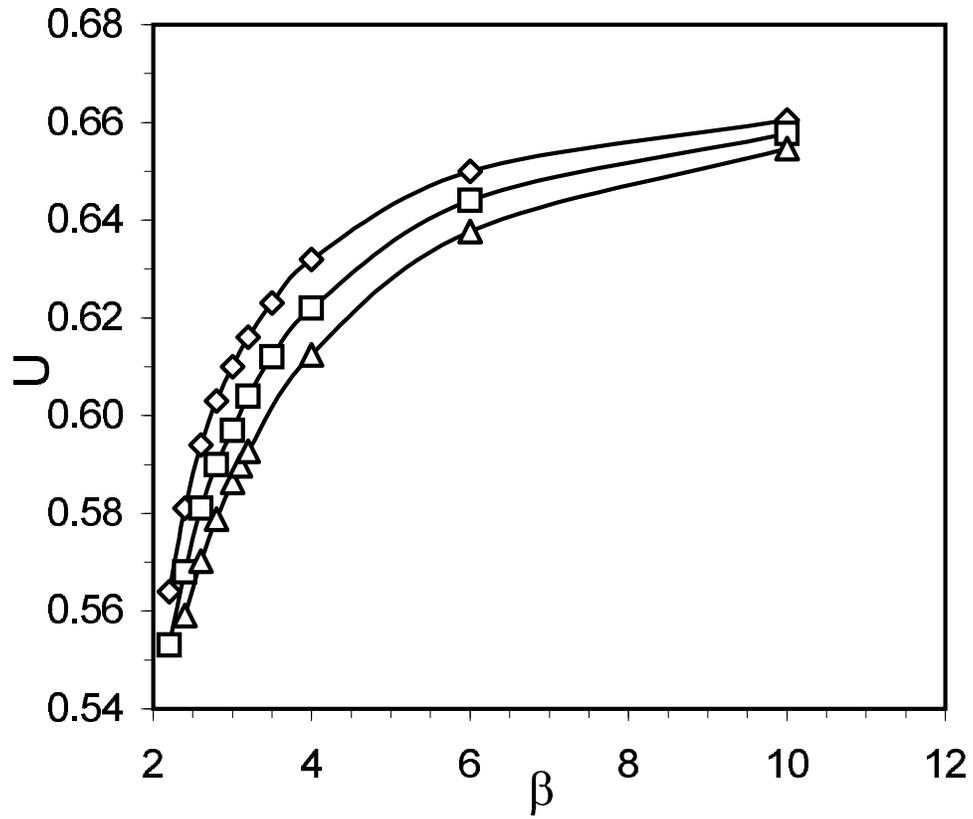}
           \caption{Fourth-order cumulant for link magnetization for the SU(2)
theory on the Wilson axis for $16^4$ (diamonds), $20^4$ (squares), 
and $24^4$ (triangles) lattices. Error bars are less than one-tenth  
the size of plotted points.}
          \label{fig10}
       \end{figure}

The deconfinement transition in SU(2) is normally considered
to be a finite-temperature transition,
one which occurs on lattices finite in at least one direction, and which disappears in the
symmetric infinite-volume limit. However, the transition observed here seems more likely
to be a bulk transition, because the order parameter is the average of a local density, and
symmetries being broken exist for any boundary
conditions. The Polyakov loop, usually invoked as the finite-temperature 
deconfinement order
parameter, and the finite temperature interpretation itself exist only for periodic 
boundary conditions.  To test this further, simulations were run with open boundary
conditions in all directions. Sample results are shown in Fig.~\ref{fig9}. Chains
were included in the measurements only if they were more than two lattice 
spacings away from any boundary and the chains themselves were terminated 
two spacings before the boundary. The symmetry breaking here appears similar to the case
of periodic boundary conditions. 
This strongly supports the idea of a bulk transition,
one which will continue to exist on the infinite lattice. It is true that the apparent critical point
does shift an unusual amount with lattice size, which is a primary motivation
for the finite-temperature interpretation of the phase transition. 
This, however, could be due in part to the small effective
size of the system. 
It seems possible the shift observed can simply be interpreted as the ordinary finite-size
shift of the bulk critical point\cite{tcp}. 
The existence of an order parameter with $\chi$ peaks allows
for a finite-size scaling analysis which should shed light on the critical exponents,
and may allow a determination of the infinite lattice critical point.  This will
probably require several much-larger lattices and higher statistics for definitive results.
Binder fourth-order cumulant crossings would cement the existence 
of a finite order transition. For the O(4) order parameter, the Binder cumulant,
defined here as 
\begin{equation}
U = 1-<m^4>/(3<m^2 >^2 ) ,
\end{equation}
varies from 1/2 in the full unbroken phase to 2/3 in the fully broken limit\cite{mdop}.
Data taken to date show Binder cumulants for the different lattice sizes
merging at weak coupling rather than crossing (Fig.~\ref{fig10}), similar to that seen
for the sub-maximal axial gauge\cite{lsb}.  Merging would be the expected
behavior for a Berezinskii-Kosterlitz-Thouless (BKT)
transition\cite{bkt}.  Three other features also favor a BKT
transition. One is the behavior of the specific heat.  The SU(2) theory shows a large
peak around $\beta = 2.2$ which does not vary significantly with lattice size.  
The BKT transition in the 2-d XY model has a specific heat curve that looks very similar
to this and is also independent of lattice size\cite{chl}.  The actual infinite order 
singularity is very soft and not visible in numerical specific heat
data. It lies at a weaker coupling, near where the specific heat {\em begins} to rise. 
Eventually a large peak, not much dependent on 
lattice size, grows far inside the strong coupling phase at the point where vortex unbinding
finally dominates. BKT transitions also exhibit an unusually large finite-lattice 
shift in apparent critical point, because the shift is logarithmic rather than a power law.
Finally, histograms for BKT
transitions near the critical point 
in the broken phase tend to show broad highly-asymmetric 
distributions that extend all the way to zero\cite{bkthistos}
not unlike those seen here, and also for the high-adjoint coupling case above.  Even though
the transition is first order there, the weak coupling phase would have to be the same as seen on
the Wilson axis, with similar properties. 

\section{Continuum Limit}
If the confining and weak-coupling phases are separated by a symmetry-breaking bulk 
phase transition, then the continuum theory, which is the weak coupling limit, will
not lie in the confining phase. It will be in a Coulomb-like phase.  This is contrary to usual 
expectations for the non-abelian case, where the confining phase is generally assumed to
extend all the way to $\beta \rightarrow \infty$ on the infinite lattice.
The situation proposed here 
would instead be comparable to the abelian case where confinement is strictly a 
strong-coupling lattice artifact, separated from the 
Coulomb phase of the
continuum theory by a phase transition.
Although in this case the continuum pure gauge non-abelian 
theory would not be confining, when
light quarks are added, one could still have chiral symmetry breaking (CSB). Physical
confinement could result as a byproduct of the CSB\cite{csb}, or through the 
closely related Gribov scenario\cite{gribov}. 

The existence of a bulk transition at finite $\beta$ could be proven analytically if
the residual gauge symmetry in the minimal axial gauge could be shown to be 
broken in some small region around $\beta = \infty$ (i.e.\ a finite region in 
$1/ \beta$ around zero).  This is because it is clearly unbroken
in the strong coupling limit ($\beta \rightarrow 0$), due to the completely random
nature of configurations there, and the lack of any energy barriers to tunneling whatsoever.
The strong coupling expansion preserves this symmetry\cite{z2h} and it undoubtedly
persists throughout the confining phase. One can see that the
symmetry is broken at $\beta = \infty$ from the following argument.  With the 1-direction links set to unity,
the 1-2, 1-3, and 1-4 plaquettes all behave as spin couplings, i.e. if the gauge links
are written as
\begin{equation}
s_0 + i\sum _{j=1}^{3} s_j \tau _j ,
\end{equation}
the 1-$k$ plaquette can be written as 
\begin{equation}
\sum _{j=0}^{3} s_{j,k,\vec{x}} s_{j,k,\vec{x}+\hat{\scriptsize{1}}},
\end{equation}
i.e. just the color dot-product of the O(4) spins.
 Here the first index is a color index
the second the link direction (2 to 4) and the third the lattice site label written 
as a spacetime vector. The symbol $\hat{1}$ represents a unit
lattice vector in the 1-direction. 
This is identical to the interaction of an O(4) spin model.
Indeed, the gauge theory can be thought of as a set of O(4) spin chains which are
linked together through ``sideways" gauge couplings carried by the remaining 2-3, 2-4 and 3-4 
plaquettes. At $\beta= \infty$ all links along a given chain must be perfectly aligned.
So the chains, even if considered in isolation, spontaneously break the symmetry. (In
other words, even one-dimensional spin chains are in the ordered phase at 
exactly $T=1/\beta =0$).  The interlinking gauge interactions cause further ordering, so 
the symmetry must remain broken in the full theory at infinite $\beta$.  Therefore
a symmetry breaking phase transition must exist - the question is whether it 
exists at finite coupling or at $\beta=\infty$ itself as in the 1-d spin chain. 
In the 1-d case, the chain disorders at 
any non-zero temperature because half of the chain can be flipped costing
energy only locally at the position of the flip; but in the current theory flipping a half-chain frustrates
sideways plaquettes not only in the flipping region, but all along the semi-infinite
flipped half-chain, with an infinite energy penalty.  This strongly suggests a behavior more 
like the higher-dimensional spin-theories, with transitions at finite couplings.  

\section{Is gauge fixing necessary?}
It is interesting to consider whether any signal of the above phase transition
can be seen in the completely unfixed theory, where Elitzur's theorem prevents
the local gauge symmetries from breaking. One can construct gauge covariant links by
multiplying both ends of any given link by a chain of links in the 1-direction, 
parallel-transporting it back to the $x_1 =0$ hypersurface. 
One can then average these covariant links
lying along chains in the 1-direction. 
These objects, when 
interpreted as O(4) vectors, are 
identical to the above link magnetizations defined in the minimal axial gauge. 
In other words, if an unfixed lattice is transformed to the minimal axial gauge, the resulting
link magnetizations will be the same as that computed from the covariant links.  This is because
the gauge transformation can be accomplished entirely by gauge transformations at sites away from 
$x_1 = 0$.  The remaining gauge transformations on $x_1 = 0$ are precisely those left unfixed in this gauge. 
The covariant links, 
however are constructed to be {\em invariant} to gauge transformations away from  $x_1 = 0$ and covariant to
those on $x_1 = 0$. The O(4) moduli of these chain-averaged covariant links are, in fact, gauge invariant. 
The signals
of spontaneous breaking used previously are all determined from the distributions of these gauge-invariant moduli
(since tunneling on a finite lattice prevents actual observation of spontaneous symmetry
breaking through symmetry-breaking vacuum expectation values anyway). 
Therefore it would appear that these phase transitions could be observed, in principle,
using gauge invariant objects in the unfixed theory.  From a practical point of view, however, it
would be expensive to compute all of the gauge-covariant links.  The minimal axial gauge 
simulation would be much faster.

There is one difference between the unfixed and minimal axial gauge simulations.  The symmetry the covariant links
are sensitive to in the unfixed simulation is a local gauge symmetry at $x_1 = 0$. In the minimal
axial gauge it is a global symmetry along each chain (independent of $x_1$) 
and local between chains. 
Elitzur's theorem prohibits the symmetry from 
breaking in the former, but not the latter.  Indeed, the direction of each chain-magnetization will drift in
the unfixed simulation due to local gauge drift at $x_1 = 0$. If these configurations are transformed to
minimal axial gauge, different configurations will map into different vacuum sectors of the gauge-fixed theory.
Thus the unfixed simulation is akin to a gauge-fixed simulation that also includes a random global/local
residual gauge transformation after each sweep. This will jump the simulation around between different vacuum sectors.
Because each vacuum sector has identical behavior except for the magnetization direction itself, all 
moments of the magnetization magnitude from which the critical behavior is derived are still identical. Only
the vacuum expectation value of the magnetization itself is erased. This should not be viewed as true tunneling.
One could artificially add global flips to an Ising model simulation, which would also erase the expectation value
of the order parameter, but nothing else about the phase transition would change. True tunneling involves 
intermediate lattice configurations 
which are partially in one vacuum sector and partially in another. If such lattices are 
energetically allowed, then tunneling can take place. It is the lack of such intermediate lattice configurations that 
is the true restriction on ergodicity which takes place at a phase transition. Another way of looking at this is that
for the truly infinite lattice, just a single configuration should be sufficient to determine any local or 
partially local quantity to arbitrary precision through spatial averaging. Ensemble averaging is, in a sense, redundant 
on the infinite lattice.  Any local condition that can exist (in the given vacuum sector) will exist somewhere
in space in each configuration.  However, for a single infinite configuration, both the gauge-fixed and unfixed
theories will exhibit vacuum expectation values of the order parameters. This demonstrates that the gauge-drift in the
former is more akin to the artificial tunneling described above than to real tunneling. Consideration of a single
infinite configuration and using spatial averaging as opposed to an ensemble average allows one to evade 
Elitzur's theorem and observe full symmetry breaking through nonlocal operators in the unfixed theory.

This work shows that over-fixing the gauge beyond the minimal
axial gauge is dangerous in the neighborhood of
the phase transition. A related question is whether it is equally dangerous within 
the weak-coupling phase itself. A gauge over-fixing that simply chooses
a particular vacuum to work in is not dangerous, if, once in that vacuum, natural
fluctuations would not, even in the absence of the extra gauge-fixing,  violate it.
However some gauge-fixings will be seen by the system as explicit symmetry
breaking on top of spontaneous symmetry breaking, which will change observables.
For instance, if one imposed the additional constraint that the averages of links along
all 1-d spin chains in the gauge-fixing direction 
lie in the unit-matrix direction (with zero components along the three
Pauli-matrix color-directions), that would simply be choosing a vacuum. Natural
fluctuations on the infinite lattice would not be able to violate this condition, so
the lattice would be ``unaware" an extra condition was being imposed.
However, if the same residual gauge freedom were used to set single links on particular chains 
equal to unity  (as in the usual maximal-tree axial gauge), then a particular
vacuum is not chosen, because all vacua have configurations in which those particular 
links are unity and others in which
they are not.
Spontaneous symmetry breaking will still select a vacuum, however, and the new fixed-links
will act as explicit symmetry breakings, affecting the natural fluctuations in that 
vacuum. The fixed link could, for instance, affect Goldstone fluctuations in its vicinity.
Thus, only additional gauge fixings which are {\em compatible} with the symmetry
breaking pattern {\em by being functions of the order parameters} are allowed. Others
will act as explicit symmetry breakings which could affect the spectrum, such as
by giving mass to Goldstone modes.  For the infinite lattice, extra gauge fixings 
on the boundary probably do not affect the bulk properties, but finite volume
or finite temperature formulations could be affected.  The implications for other
popular gauges such as Landau and Coulomb are not immediately clear, and are worth
investigating.

\section{Conclusion}
   The concept of symmetry breaking in gauge theories has been a confusing one. On one hand, 
Elitzur's theorem prohibits spontaneous symmetry breaking of local symmetries.  On the other,
the standard model relies on spontaneous breaking of the gauge symmetry to initiate the Higgs mechanism.
One way to reconcile these is with partial gauge fixing.  If one fixes the gauge enough to
make it global in at least one direction, then Elitzur's theorem no longer applies, allowing the remaining
residual symmetries to break spontaneously, if this is energetically favored. This is apparently the case in continuum QED,
where it is even possible to interpret the photon as a Goldstone boson.  In this paper, the possible
breaking of such residual symmetries was explored in pure lattice gauge theories. It was found that the residual
gauge symmetries
are spontaneously broken at weak coupling in both the 3-d Z2 and 4-d SU(2) theories, and that these are
separated from the strong-coupling confining region by a phase transition.  This may be true for most if not all
gauge theories in the minimal axial gauge, where the symmetry is fixed in only one direction, leaving it
still local in the others. Since it is presumably dangerous to add explicit symmetry breaking on top
of spontaneous symmetry breaking, gauges more restrictive than this could introduce unphysical effects.
Thus the partial spontaneous breaking of the gauge symmetry itself appears to violate the concept of gauge-equivalence.
This requires further investigation.

On the lattice, gauge theories have a resemblance to magnetic spin models. 
In the axial gauge, this resemblance
is strengthened. In two dimensions gauge and spin theories are equivalent, 
and in three they are sometimes related to each other by duality transformations.
In four dimensions half of the plaquette interactions become spin interactions in the axial gauge, 
leading to a picture of the
gauge theory being a system of interacting one-dimensional spin chains. The critical behavior in
the minimal axial gauge is
especially interesting because it has some features of a four-dimensional system but some of a 
one-dimensional system. This is due to the order parameters for each of the many broken symmetries being 
averaged only
over each associated 1-d spin chain.  When viewed as magnetic systems, it is not surprising to find the spins
to be magnetized at weak coupling (low effective 4-d temperature). The disparate behaviors exhibited by
different gauge theories may be related to the mode of symmetry breaking, either Nambu-Goldstone, Higgs, or BKT.
Clearly, there are many aspects of this phenomenon to be explored.

\newpage
\newpage
\newpage
\newpage
\newpage
\newpage
\newpage
\newpage
\newpage
\newpage

\end{document}